\documentclass[12pt]{article}
\usepackage{amsfonts}
\usepackage{epsfig,amssymb,euscript}
\usepackage{amsmath,amscd}
\usepackage{slashed}

\numberwithin{equation}{section}
\def\appendix#1{\addtocounter{section}{1}\setcounter{equation}{0}
\renewcommand{\thesection}{\Alph{section}}
\section*{Appendix \thesection\protect\indent \parbox[t]{11.15cm}{#1}}
\addcontentsline{toc}{section}{Appendix \thesection\ \ \ #1}}

\begin{document}

\begin{titlepage}
\begin{center}

\vspace*{-1.0cm}

\hfill DMUS-MP-20/04
\vspace{2.0cm}

\renewcommand{\thefootnote}{\fnsymbol{footnote}}
{\Large{\bf Neutral Signature Gauged Supergravity Solutions}}
\vskip1cm
\vskip 1.3cm
J. B. Gutowski$^1$  and W. A. Sabra$^2$
\vskip 1cm
{\small{\it
$^1$Department of Mathematics,
University of Surrey \\
Guildford, GU2 7XH, UK.\\}}
\vskip .6cm {\small{\it
$^2$ Physics Department, 
American University of Beirut\\ Lebanon  \\}}
\end{center}
\bigskip
\begin{center}
{\bf Abstract}
\end{center}
We classify all supersymmetric solutions of minimal $D=4$ gauged supergravity with $(2,2)$ signature and a positive cosmological constant which admit exactly one Killing spinor. This classification produces a geometric structure which is more general than that found for previous classifications of $N=2$ supersymmetric solutions of this theory.
We illustrate how the $N=2$ solutions which consist of a fibration over a 3-dimensional Lorentzian Gauduchon-Tod base space can be written in terms of this more generic geometric structure.

\end{titlepage}

\section{Introduction}
Substantial progress has been made in the classification and understanding of solutions admitting supersymmetry in four-dimensional supergravity theories. This  was initiated by the work of Tod \cite{Tod:1983pm, Tod:1995jf} which dealt with the classification of all solutions admitting Killing spinors in Lorentzian four-dimensional $N=2$ supergravity theory. Later, systematic classifications were performed for the Lorentzian four-dimensional gauged $N=2$ supergravity theory in \cite{Caldarelli:2003pb, Cacciatori:2008ek, Klemm:2009uw}. This subsequent work relied on the use of Fierz identities. Other work classifying supersymmetric four-dimensional solutions coupled to hypermultiplets  was done in \cite{Meessen:2010fh}. 
    
     In recent years, spinorial geometry techniques have proven to be particularly 
useful for the analysis and classification of supersymmetric solutions. This method relies on expressing spinors in terms of differential forms \cite{forms1, forms2}. In using this method to classify supergravity solutions, one exploits the gauge symmetries of the theory to find simple canonical forms for the Killing spinors. Those canonical forms are then employed to find a linear system which facilitates the finding of solutions for the Killing spinor equations. The spinorial geometry techniques were first employed in the classification of solutions with Killing spinors in $D=11$ supergravity, heterotic and type II supergravity theories in \cite{Gillard:2004xq, Gran:2005wu, Gran:2005wn, Gran:2005kg, Gran:2005wf}. Also more work in the classification of four-dimensional solutions using spinorial geometry was performed in \cite{Gutowski:2009vb, Gutowski:2018shj}. For more details on the spinorial geometry applications to the classification of supersymmetric solutions, we refer the reader to the review \cite{Gran:2018ijr}.
    The classification of solutions with Killing spinors was extended to the cases of four-dimensional Euclidean $N=2$ supergravity in \cite{euclidean1, euclidean2, euclidean3}, by making use of the 2-component spinor formalism and the spinorial geometry techniques. In this analysis, interesting relations of supersymmetric solutions to Einstein-Weyl structures, and the $SU(\infty)$ Toda equation were discovered.

The case of geometries with neutral signature, i.e. signature $(+,+,-,-)$, is of particular
interest, partly because such the properties of such solutions have not been extensively explored
in the literature, at least in comparison to the Euclidean and $(-,+,+,+)$ Lorentizian cases.
The analysis of neutral signature solutions admitting parallel spinors in four-dimensional gravity has been performed in \cite{bryant, md, mdp}, where solutions admitting null-K\"ahler structures were obtained. Useful and extensive details on the neutral signature theory as well as some of its solutions can be found in \cite{Barrett:1993yn}. 
    
    A systematic classification of solutions in four-dimensional Einstein-Maxwell theory with or without cosmological constant in neutral signature was considered in \cite{Klemm:2015mga}. In a more recent paper \cite{Gutowski:2019hnl}, we have used spinorial geometry techniques to classify supersymmetric solutions of the the minimal four-dimensional Einstein-Maxwell with neutral signature and without a cosmological constant. Such a theory can be thought of as a truncation of the  $N=2$ supergravity theory obtained from Hull's M theory \cite{ch} via a reduction on $CY_3 \times S^1$ \cite{Sabra:2017xvx}. In the analysis of \cite{Gutowski:2019hnl}, two orbits for the Majorana Killing spinors were found using appropriately chosen $Spin(2,2)$ gauge transformations. The solutions for the orbit represented by a chiral Killing spinor were found to correspond to a sub-class of the solutions found \cite{bryant, md, mdp}. Moreover, a novel geometric structure was discovered for the solutions corresponding to the second non-chiral orbit. 
    
   Our current work is concerned with the classification of all solutions admitting one Killing spinor in minimal $D=4$ gauged supergravity with $(2,2)$ signature and a positive cosmological constant. This  however does not simply reduce to a subclass of the solutions considered previously in \cite{Klemm:2015mga}, because a class of solutions corresponding to the special case for which there is a single Majorana Killing spinor was omitted from the
classification in that work. In fact, all of the solutions with positive cosmological constant
classified in \cite{Klemm:2015mga} have $N=2$ supersymmetry. The purpose of this work
is to understand the geometric properties of solutions with minimal $N=1$ supersymmetry,
associated with a Majorana Killing spinor, and to find examples of such solutions.

We plan our paper as follows.  In Section two, we introduce the Killing spinor equation of our theory as well as a summary of conventions and results of \cite{Gutowski:2019hnl} which are relvant to our subsequent analysis. Relations involving two $Spin(2,2)$ invariant spinor bilinears are presented and it is demonstrated that only one of the two canonical Killing spinors obtained in \cite{Gutowski:2019hnl} survives in the gauged theory. 
Section three contains the analysis of the Killing spinor equations for the remaining non-chiral orbit and the geometric conditions arising from this analysis. In section four, we present specific examples of our general solutions corresponding to solutions which are fibrations over a Lorentzian Gauduchon-Tod space as well the specific example corresponding to the (2,2) analogue of Kastor-Traschen solution. Self-dual solutions are explored in section five and we conclude in section six. Appendix A contains the linear system corresponding to the non-chiral Killing spinor.

\section{Majorana Spinor Orbits}

In this section, we introduce the Killing spinor equation which
we shall analyse, and summarize a number of results concerning spinor conventions from \cite{Gutowski:2019hnl} which will be particularly useful. The Killing spinor equation (KSE)  is given by
\begin{eqnarray}
\label{kse1}
D_\mu \epsilon \equiv \nabla_\mu \epsilon- {1 \over 4}{\slashed F} \Gamma_\mu \epsilon 
-{1 \over 2 \ell} \Gamma_\mu \epsilon -{1 \over \ell}A_\mu \epsilon
=0
\end{eqnarray}
where $F$ is the Maxwell field strength, $F=dA$, which satisfies
\begin{eqnarray}
dF=0, \qquad d \star F=0 \ .
\end{eqnarray}
For spinors, we adopt the same conventions as introduced in
\cite{Gutowski:2019hnl}. In particular, there exists
a charge conjugation operator $C*$,  $[C*,\Gamma_\mu]=0$,
and hence if $\epsilon$ satisfies ({\ref{kse1}}) then so does $C*\epsilon$. In this paper, we shall concentrate in particular on Majorana Killing spinors $\epsilon$ which satisfy $C*\epsilon=\epsilon$. We note that this class of solutions
was omitted from the classification constructed in
\cite{Klemm:2015mga}. In particular, supersymmetric
solutions of the minimal $D=4$ gauged supergravity in neutral
signature with a positive cosmological constant must admit
spinors satisfying ({\ref{kse1}}), and for a solution
to preserve the minimal possible $N=1$ supersymmetry, then
the Killing spinor must be Majorana, or otherwise a second
linearly independent Killing spinor can be constructed by
making use of $C*$. Hence, we shall classify the
minimal $N=1$ solutions of ({\ref{kse1}}). In contrast, all
of the supersymmetric solutions of this theory classified
in \cite{Klemm:2015mga} preserved $N=2$ supersymmetry, and are
therefore special cases of a more generic structure corresponding to the $N=1$ case, which we present here.

As observed in \cite{Gutowski:2019hnl}, a generic Majorana spinor can be put into one of two possible canonical forms
using gauge transformations. In the first canonical form, the spinor is chiral, whereas in the second canonical form, it is not.
The canonical forms can be further characterized in terms of certain $Spin(2,2)$ invariant spinor bilinears. The spinor bilinears are 1-forms
and 2-forms $W$ and $\chi$, given by
\begin{eqnarray}
\label{bilin}
W_\mu = i{\cal{B}}(\epsilon, \Gamma_5 \Gamma_\mu \epsilon),
\qquad
\chi_{\mu \nu} = i {\cal{B}}(\epsilon, \Gamma_5 \Gamma_{\mu \nu} \epsilon) \ ,
\end{eqnarray}
where the inner product ${\cal{B}}$ satisfies
\begin{eqnarray}
{\cal{B}}(\epsilon,\eta)=-{\cal{B}}(\eta,\epsilon), \qquad
{\cal{B}}(\epsilon, \gamma_\mu \eta)= {\cal{B}}(\gamma_\mu \epsilon,  \eta)
\end{eqnarray}
for Majorana $\epsilon, \eta$. Explicit expressions
for $W$ and $\chi$ for the two different canonical
Majorana spinor orbits have been evaluated in \cite{Gutowski:2019hnl}. Here it suffices to note that in the chiral Majorana orbit,
the 1-form vanishes $W=0$, but $\chi \neq 0$. In the non-chiral 
Majorana orbit, both $W$ and $\chi$ are non-vanishing, and satisfy
\begin{eqnarray}
W^2=0, \qquad \chi = W \wedge \theta, \qquad W \cdot \theta =0, \qquad \theta^2=1 \ .
\end{eqnarray}

The Killing spinor equation ({\ref{kse1}}) implies the following conditions
\begin{eqnarray}
\label{cc3}
\nabla_\nu W_\mu = {1 \over 2}\eta_{\mu \nu} F_{\lambda_1 \lambda_2}\chi^{\lambda_1 \lambda_2}+ F_{\nu \lambda}\chi^\lambda{}_\mu + F_{\mu \lambda} \chi^\lambda{}_\nu
+{2 \over \ell} A_\nu W_\mu +{1 \over \ell} \chi_{\mu \nu}
\end{eqnarray}
and
\begin{eqnarray}
\label{cc4}
\nabla_\sigma \chi_{\mu \nu}&=& F_{\sigma \mu}W_\nu - F_{\sigma \nu} W_\mu
- F_{\mu \nu}W_\sigma +\eta_{\sigma \mu} (i_W F)_\nu - \eta_{\sigma \nu}(i_W F)_\mu 
\nonumber \\
&+&{2 \over \ell} A_\sigma \chi_{\mu \nu} -{1 \over \ell} \eta_{\mu \sigma}
W_\nu +{1 \over \ell} \eta_{\nu \sigma} W_\mu \ .
\end{eqnarray}

Hence, for the chiral orbit for which $W=0$, $\chi \neq 0$,
a contradiction is immediately obtained by taking the antisymmetric part of ({\ref{cc3}}). There are therefore no
Majorana Killing spinors in this orbit for the gauged theory.
It remains to analyse the Killing spinor equations for the remaining non-chiral Majorana orbit.

\section{Analysis of the Killing Spinor Equation}

In this section, we analyse the Killing spinor equation
({\ref{kse1}}), when the Majorana spinor $\epsilon$ is in the non-chiral orbit, and we derive the necessary and sufficient conditions on the geometry and the Maxwell field strength.

In order to perform the analysis, we shall use spinorial geometry techniques which were originally introduced
for the analysis of $D=10$ and $D=11$ supersymmetric supergravity 
solutions \cite{Gillard:2004xq, Gran:2005wu, Gran:2005wn, Gran:2005kg, Gran:2005wf}. Following the same analysis as
set out in \cite{Gutowski:2019hnl}, the Majorana 
spinor in the non-chiral orbit can be written
in a specific basis as $\epsilon=1+e_{12}+e_1+e_2$. The linear system relating the spin connection and components of the gauge potential and field strength is listed in Appendix A.
In particular, it can be shown that the linear system 
({\ref{linsys}) is equivalent to the conditions
({\ref{cc3}}) and ({\ref{cc4}}), which are therefore necessary
and sufficient conditions for supersymmetry. We shall therefore
analyse the conditions ({\ref{cc3}}) and ({\ref{cc4}}).

To proceed, contract ({\ref{cc3}}) with $W^\nu$ to obtain
\begin{eqnarray}
\nabla_W W = {2 \over \ell} (i_W A) W \ .
\end{eqnarray}
By making an appropriately chosen $U(1) \times Spin(2,2)$ gauge transformation which leaves the Killing spinor invariant, we may without loss of generality work in a gauge for which 
\begin{eqnarray}
i_W A =0
\end{eqnarray}
and in this gauge
\begin{eqnarray}
\label{geo1}
\nabla_W W=0 \ .
\end{eqnarray}
Furthermore, taking the trace of ({\ref{cc3}}) implies that
\begin{eqnarray}
\label{coclos}
\nabla_\mu W^\mu=0 \ .
\end{eqnarray}
The antisymmetric part of ({\ref{cc3}}) implies that
\begin{eqnarray}
\label{clos1}
dW= {2 \over \ell}(A+\theta)\wedge W
\end{eqnarray}
and hence on taking the exterior derivative, one finds
\begin{eqnarray}
\label{auxt1}
W \wedge (F+d \theta)=0 \ .
\end{eqnarray}
The condition ({\ref{cc4}}) also implies that
\begin{eqnarray}
\label{auxt2}
(i_W F)_\nu = \nabla^\lambda \chi_{\lambda \nu} +\bigg({3 \over \ell}+{2 \over \ell}\theta^\mu A_\mu \bigg) W_\nu \ .
\end{eqnarray}
On substituting ({\ref{auxt1}}) and ({\ref{auxt2}}) into ({\ref{cc4}}) one finds
that ({\ref{cc4}}) is equivalent to
\begin{eqnarray}
W_\sigma F_{\mu \nu} &=& -{1 \over 2} \nabla_\sigma \chi_{\mu \nu} -{1 \over 2} (W \wedge d \theta)_{\sigma \mu \nu}
+ \eta_{\sigma \mu} \bigg({1 \over 2} \nabla^\lambda \chi_{\lambda \nu}
+{1 \over \ell}(1+A^\lambda W_\lambda) W_\nu \bigg)
\nonumber \\
&-& \eta_{\sigma \nu} \bigg({1 \over 2} \nabla^\lambda \chi_{\lambda \mu}
+{1 \over \ell}(1+A^\lambda W_\lambda) W_\mu \bigg)
+{1 \over \ell}A_\sigma \chi_{\mu \nu} \ .
\end{eqnarray}
This condition can be used to eliminate $F$ entirely from ({\ref{cc3}}).
The resulting conditions obtained from ({\ref{cc3}}) are ({\ref{clos1}}), together with
\begin{eqnarray}
\label{auxt3}
\nabla_W \theta = \bigg({\nabla^\lambda \theta_\lambda} -{2 \over \ell} \bigg) W \ .
\end{eqnarray}
To continue, consider ({\ref{clos1}}), which implies that 
\begin{eqnarray}
W \wedge dW =0
\end{eqnarray}
and hence there exists a local co-ordinate $u$ and a function $H$, not identically zero, such that
\begin{eqnarray}
W= H du \ .
\end{eqnarray}
On substituting this expression back into ({\ref{clos1}}), one obtains
\begin{eqnarray}
A+\theta-{\ell \over 2} H^{-1} dH = {\cal{G}} W
\end{eqnarray}
for some function ${\cal{G}}$. There is a freedom to redefine $\theta$ as
${\hat{\theta}}=\theta-{\cal{G}} W$, as well as making an appropriately chosen $U(1) \times Spin(2,2)$ gauge transformation which leaves the Killing spinor invariant
and preserves the condition $i_W A=0$. Then without loss of generality, we may take
\begin{eqnarray}
A=- \theta
\end{eqnarray}
and in this gauge
\begin{eqnarray}
dW=0
\end{eqnarray}
and furthermore ({\ref{auxt2}}) implies that
\begin{eqnarray}
\nabla^\lambda \theta_\lambda = {3 \over \ell}
\end{eqnarray}
and so ({\ref{auxt3}}) simplifies further to
\begin{eqnarray}
\label{geom1}
\nabla_W \theta = {1 \over \ell} W \ .
\end{eqnarray}

Next, substitute the condition $A=-\theta$, $F=-d \theta$ back into the equations
({\ref{cc3}}) and ({\ref{cc4}}) to obtain
\begin{eqnarray}
\label{cc3a}
\nabla_\nu W_\mu +{2 \over \ell}W_{(\mu}\theta_{\nu)}=\eta_{\mu \nu} \theta^\lambda W^\rho (d \theta)_{\lambda \rho}
+2 (i_W d \theta)_{(\nu} \theta_{\mu)}-2(i_\theta d \theta)_{(\nu}W_{\mu)}
\end{eqnarray}
and
\begin{eqnarray}
\label{cc4a}
{1 \over 2}W_\sigma (d \theta)_{\mu \nu} -W_{[\mu}\nabla_{\nu]} \theta_\sigma
- \bigg(-(i_\theta d \theta)_\sigma W_{[\mu}+\theta_{\sigma} (i_W d \theta)_{[\mu}
-W_\sigma (i_\theta d \theta)_{[\mu} \bigg) \theta_{\nu]}
\nonumber \\
-\eta_{\sigma [\mu} \bigg(\theta_{\nu]} (d \theta)_{\lambda \rho} \theta^\lambda W^\rho
+(i_W d \theta)_{\nu]} +{1 \over \ell} W_{\nu]}\bigg) -{1 \over \ell} \theta_\sigma W_{[\mu} \theta_{\nu]}=0 \ .
\end{eqnarray}
In order to obtain the conditions on $\nabla \theta$ obtained from ({\ref{cc4}}) it is
most straightforward to set
\begin{eqnarray}
\Psi_{\mu \nu} = \nabla_\mu \theta_\nu+{1 \over \ell} \theta_\mu \theta_\nu -{1 \over \ell} \eta_{\mu \nu} \ .
\end{eqnarray}
On substituting this expression into ({\ref{cc4a}}), the resulting expression
is identical to ({\ref{cc4a}}) excluding the $\ell^{-1}$ terms, on replacing
$\nabla_\mu \theta_\nu$ with $\Psi_{\mu \nu}$. Furthermore, $\Psi_{\mu \nu}$ satisfies
the conditions $\Phi_\mu{}^\mu=0$ and $W^\mu \Psi_{\mu \nu}=0$, which are the conditions
satisfied by $\nabla_\mu \theta_\nu$ in the case of the ungauged theory. The corresponding
geometric condition obtained from ({\ref{cc4a}}) is therefore obtained from the corresponding
condition in the analysis of the ungauged theory, replacing $\nabla_\mu \theta_\nu$ with
$\Psi_{\mu \nu}$, which is
\begin{eqnarray}
\label{extrageo1}
\nabla_\tau \theta = \star (\theta \wedge d \theta) +{1 \over \ell} \tau \ .
\end{eqnarray}
A similar analysis of ({\ref{cc3a}}), taking the condition obtained on $\nabla W$
obtained in the ungauged theory, and replacing $\nabla_\nu W_\mu$ with
$\nabla_\nu W_\mu +{2 \over \ell}W_{(\mu}\theta_{\nu)}$, implies that
\begin{eqnarray}
\label{extrageo2}
\nabla_V W = \star(\tau \wedge d \theta) - i_\theta d \theta -{1 \over \ell} \theta
\end{eqnarray}
and
\begin{eqnarray}
\label{extrageo3}
\nabla_\tau W = \star (W \wedge d\theta)
\end{eqnarray}
where we adopt a frame $\{ V, W, \tau, \theta \}$, with respect to which the metric is
\begin{eqnarray}
\label{newframe}
ds^2=2VW+\theta^2-\tau^2
\end{eqnarray}
with volume form ${\rm dvol}=W \wedge V \wedge \tau \wedge \theta$.

In order to analyse the conditions obtained from the gauge field equations, 
we note that 
\begin{eqnarray}
\star \chi = -W \wedge \tau
\end{eqnarray}
and that the condition ({\ref{cc4}}) can be rewritten as
\begin{eqnarray}
\nabla_\sigma \star \chi_{\mu \nu} &=& -W_\sigma \star F_{\mu \nu}
+W_\mu \star F_{\nu \sigma}-W_\nu \star F_{\mu \sigma}
+ \eta_{\sigma \mu} (i_W \star F)_\nu - \eta_{\sigma \nu} (i_W \star F)_\mu
\nonumber \\
&+&{2 \over \ell} A_\sigma \star \chi_{\mu \nu} -{1 \over \ell} (\star W)_{\sigma \mu \nu}
\nonumber \\
\end{eqnarray}
and hence
\begin{eqnarray}
d \star \chi = W \wedge \bigg(\star F +{1 \over \ell} \theta \wedge \tau \bigg) \ .
\end{eqnarray}
This implies that
\begin{eqnarray}
W \wedge \bigg(\star F - d \tau +{1 \over \ell} \theta \wedge \tau \bigg)=0
\end{eqnarray}
and therefore there exists a 1-form $\xi$ such that
\begin{eqnarray}
\star F = d \tau -{1 \over \ell} \theta \wedge \tau +W \wedge \xi
\end{eqnarray}
and hence
\begin{eqnarray}
\label{extrageo4}
d \tau = {1 \over \ell} \theta \wedge \tau -W \wedge \xi - \star d \theta \ .
\end{eqnarray}
The gauge field equations therefore imply that
\begin{eqnarray}
W \wedge d \xi +{1 \over \ell} d \big(\theta \wedge \tau \big)=0 \ .
\end{eqnarray}

Furthermore, further simplification can be made by considering the 
basis transformation
\begin{eqnarray}
\label{basist1}
V'=V-\beta \tau +{1 \over 2} \beta^2 W, \qquad \tau'=\tau-\beta W
\end{eqnarray}
for arbitrary function $\beta$.
Under this transformation, all the previous conditions on the geometry are invariant,
with $\xi$ replaced with
\begin{eqnarray}
\xi'= \xi +{\beta \over \ell} \theta -d \beta -kW
\end{eqnarray}
for arbitrary function $k$. In particular, $\beta$ can be chosen such that 
$i_W \xi' =0$, and $k$ can also be chosen such that $i_V' \xi'=0$. Adopting this basis, and dropping the prime, we take without loss
of generality $i_W \xi = i_V \xi =0$, and hence there exist functions $f_1, f_2$ such that
\begin{eqnarray}
\xi = f_1 \theta + f_2 \tau \ .
\end{eqnarray}
Then it is straightforward to prove, on making use of ({\ref{extrageo4}}), that the condition ({\ref{extrageo3}}) is equivalent
to 
\begin{eqnarray}
\nabla_W \tau =0
\end{eqnarray}
and hence it follows that
\begin{eqnarray}
\nabla_W V =-{1 \over \ell} \theta \ .
\end{eqnarray}

\subsection{Summary of Geometric Conditions}
\label{conds}

We briefly summarize the necessary and sufficient conditions for supersymmetry obtained.
The frame is $\{ V, W, \tau, \theta \}$, with respect to which the metric is
\begin{eqnarray}
ds^2=2VW+\theta^2-\tau^2
\end{eqnarray}
with volume form ${\rm dvol}=W \wedge V \wedge \tau \wedge \theta$. The conditions on the geometry are then given by
\begin{eqnarray}
\label{gsum1}
dW=0, \quad \nabla_W W=\nabla_W \tau=0, \quad \nabla_W V=-{1 \over \ell} \theta, \quad
\nabla_W \theta = {1 \over \ell} W
\end{eqnarray}
together with
\begin{eqnarray}
\label{gsum2}
\star d \star \theta = -{3 \over \ell}, \quad \nabla_\tau \theta = \star (\theta \wedge d \theta) +{1 \over \ell} \tau, \quad \nabla_V W = \star(\tau \wedge d \theta) - i_\theta d \theta -{1 \over \ell} \theta
\end{eqnarray}
and there exists a 1-form $\xi$, satisfying $i_W \xi =i_V \xi=0$ such that
\begin{eqnarray}
\label{dtxx}
d \tau = {1 \over \ell} \theta \wedge \tau -W \wedge \xi - \star d \theta \ .
\end{eqnarray}
The gauge potential is $A=-\theta$, $F=-d \theta$. We remark that the condition
$d \star W=0$ has been omitted from the above conditions, this is because $d \star W=0$ is implied by the above.

There are two natural $Spin(2,2)$-invariant 2-form bilinears, which are the self-dual and anti-self-dual parts of $\chi$
\begin{eqnarray}
\omega^\pm = W \wedge (\theta \pm \tau)
\end{eqnarray}
with associated commuting nilpotent endomorphisms $J^\pm$,
\begin{eqnarray}
\omega^\pm (X,Y)= g(X, J^\pm Y)
\end{eqnarray}
for all vector fields $X, Y$. $J^\pm$ satisfy $(J^\pm)^2=0$, and $J^\pm$ are rank 2,
with ${\rm Im} J^\pm ={\rm Ker} J^\pm$. Hence, in order for $J^\pm$ to be integrable, 
we require that ${\rm Ker} J^\pm$ be closed under the Lie bracket, which is equivalent to
requiring that
\begin{eqnarray}
J^\pm [J^\pm X, J^\pm Y]=0
\end{eqnarray}
for all vector fields $X, Y$, or equivalently,
\begin{eqnarray}
\label{npident}
(J^\pm)^\mu{}_\lambda (J^\pm)^\rho{}_\alpha \nabla_\rho (J^\pm)^\lambda{}_\beta
- (J^\pm)^\mu{}_\lambda (J^\pm)^\rho{}_\beta \nabla_\rho (J^\pm)^\lambda{}_\alpha =0 \ .
\end{eqnarray}
The conditions ({\ref{gsum1}}), ({\ref{gsum2}}) and ({\ref{dtxx}}) imply that 
({\ref{npident}}) can be rewritten as
\begin{eqnarray}
\label{icond}
\omega^\mp \wedge (d \theta)^\mp =0 
\end{eqnarray}
where $(d \theta)^{\mp}$ are the anti-self-dual/self-dual parts of $d \theta$.
However, the integrability condition ({\ref{icond}}) does not hold automatically as
a consequence of ({\ref{gsum1}}), ({\ref{gsum2}}) and ({\ref{dtxx}}); though, if
$F$ is self-dual, then $(d \theta)^-=0$, and so $J^+$ is integrable.

\section{Example: Lorentzian Gauduchon-Tod Solutions}

A supersymmetric solution for which the geometry is a fibration
over a Lorentzian Gauduchon-Tod space was found in \cite{Klemm:2015mga} which preserves $N=2$ supersymmetry. In this section, we demonstrate how this solution can be written in terms
of the conditions given in Section {\ref{conds}} which
all $N=1$ supersymmetric solutions must satisfy. The metric is given by{\footnote{We remark that the metric ({\ref{gtmet}}) differs from that given in \cite{Klemm:2015mga} by an overall minus sign, which arises from a difference in conventions for the definition of the gamma matrices appearing in the KSE.}}
\begin{eqnarray}
\label{gtmet}
ds^2 = f(dt+\omega)^2+ f^{-1} ds_3{}^2
\end{eqnarray}
where
\begin{eqnarray}
f= {1 \over \ell^{-2} t^2-1}
\end{eqnarray}
and $ds_3^2$ is the $t$-independent metric on the Lorentzian Gaudochon-Tod space. A basis $\{ V^1, V^2, V^3 \}$ for
 the Lorentzian Gaudochon-Tod space is chosen such that
\begin{eqnarray}
ds_3{}^2 = -(V^1)^2-(V^2)^2+(V^3)^2
\end{eqnarray}
where
\begin{eqnarray}
\label{lgta}
d V^i = {2 \over \ell} {\cal{H}} \wedge V^i -{2 \over \ell} \star_3 V^i
\end{eqnarray}
and ${\cal{H}}$ is a $t$-independent 1-form on the Lorentzian Gauduchon-Tod space which satisfies
\begin{eqnarray}
\label{cdh}
d {\cal{H}} = {2 \over \ell} \star_3 {\cal{H}} \ .
\end{eqnarray}
In addition, 
\begin{eqnarray}
\omega = {2 \over \ell} t {\cal{H}}+\phi
\end{eqnarray}
where $\phi$ is a $t$-independent 1-form on the Lorentzian Gauduchon-Tod space satisfying
\begin{eqnarray}
\label{dccond}
d \phi ={2 \over \ell} \phi \wedge {\cal{H}} +{2 \over \ell} \star_3 \phi \ .
\end{eqnarray}
The volume form on the Lorentzian Gauduchon-Tod space is
\begin{eqnarray}
{\rm dvol}_3 = V^1 \wedge V^2 \wedge V^3
\end{eqnarray}
and the gauge potential is given by
\begin{eqnarray}
\label{gtpot}
A = - \ell^{-1} ft (dt+\omega) +{\cal{H}} \ .
\end{eqnarray}

In order to rewrite this solution in terms of the conditions
set out in Section {\ref{conds}}, it is first useful to work with local co-ordinates on the Lorentzian Gauduchon-Tod space. To obtain these, note that the conditions ({\ref{lgta}}) imply that
\begin{eqnarray}
(V^1 - V^3) \wedge d (V^1 - V^3)=0, \qquad (V^1+V^3) \wedge d (V^1+V^3) =0
\end{eqnarray}
and hence there exist local co-ordinates $p, q$, and $t$-independent functions
$h, x$ such that
\begin{eqnarray}
V^1-V^3 = h dp, \qquad V^1+V^3 = h e^x dq
\end{eqnarray}
and hence
\begin{eqnarray}
\label{lgtb}
V^1 = {1 \over 2} h dp +{1 \over 2} e^x h dq,
\qquad V^3 = -{1 \over 2} h dp +{1 \over 2} e^x h dq \ .
\end{eqnarray}
On substituting these expressions for $V^1$ and $V^3$ back into  ({\ref{lgta}}) for $i=1,3$, one further finds
\begin{eqnarray}
\label{lgtq}
V^2= {\ell \over 4} dx +{1 \over 2} e^x g_2 h dq -{1 \over 2}
g_1 h dp
\end{eqnarray}
and
\begin{eqnarray}
\label{lgtd}
{\cal{H}}= {\ell \over 4} dx +{\ell \over 2} h^{-1} dh +{1 \over 2} e^x g_2 h dq +{1 \over 2} g_1 h dp \ .
\end{eqnarray}
Then, substituting these expressions into the final condition
obtained from ({\ref{lgta}}) for $i=2$, one further finds
that the functions $h, g_1, g_2$ must satisfy
\begin{eqnarray}
\label{lgte}
e^x g_2 + e^x \partial_x g_2 +{\ell \over 2} h^{-2} \partial_q h &=&0
\nonumber \\
g_1 -\partial_x g_1 +{\ell \over 2} h^{-1} \partial_p h &=&0
\nonumber \\
2 \ell^{-1} h e^x(1-g_1 g_2) + e^x \partial_p g_1 + \partial_q g_1 &=&0 \ .
\end{eqnarray}
Hence, the Lorentzian Gauduchon-Tod space has local co-ordinates $\{ x,p,q \}$, and basis elements $V^i$ given by
({\ref{lgtb}}) and ({\ref{lgtq}}) given in terms of
functions $h=h(x,p,q)$, $g_1=g_1(x,p,q)$, $g_2=g_2(x,p,q)$,
which must satisfy the conditions ({\ref{lgte}}), and
${\cal{H}}$ is given by ({\ref{lgtd}}). With these conventions,
\begin{eqnarray}
\star_3 dp &=& \bigg({\ell \over 4} dx +{1 \over 2} e^x g_2 h dq \bigg) \wedge dp
\nonumber \\
\star_3 dq &=& -\bigg({\ell \over 4} dx -{1 \over 2} g_1 h dp \bigg) \wedge dq
\nonumber \\
\star_3 dx &=& 2 \ell^{-1} e^x h^2 (1-g_1 g_2) dp \wedge dq
+{1 \over 2} dx \wedge \bigg( e^x g_2 h dq + g_1 h dp \bigg)
\end{eqnarray}
and the condition ({\ref{cdh}}) is equivalent to
\begin{eqnarray}
dh \wedge V^2 + d(g_1 h) \wedge (V^1-V^3)
-4 \ell^{-1} h V^1 \wedge V^3 -4 \ell^{-1} h g_1 V^2 \wedge (V^1-V^3)= \star_3 dh
\nonumber \\
\end{eqnarray}
which in turn can be rewritten as
\begin{eqnarray}
\label{rxextra}
2 \ell^{-1} e^x(1-g_1 g_2) \partial_x h -e^x g_2 h^{-1} \partial_p h + g_1 h^{-1} \partial_q h + \partial_q g_1 -2 \ell^{-1} e^x h(g_1 g_2 -1) =0 \ .
\nonumber \\
\end{eqnarray}
This condition is however implied by the conditions listed in
({\ref{lgte}}). 

Having obtained the conditions on the Lorentzian Gauduchon-Tod structure in these local co-ordinates, the 4-dimensional solution can be further simplified by setting 
\begin{eqnarray}
t=h^{-1} u, \qquad \phi = h^{-1} \psi
\end{eqnarray}
so that the metric is
\begin{eqnarray}
\label{gtmet2}
ds^2 &=& {1 \over \ell^{-2} u^2-h^2}
\bigg(du+u({1 \over 2}dx +\ell^{-1} e^x g_2 h dq+\ell^{-1}g_1 h dp) + \psi\bigg)^2
\nonumber \\
&+&(\ell^{-2} u^2-h^2) \bigg(-e^x dp dq -\big({\ell \over 4} h^{-1} dx +{1 \over 2} e^x g_2 dq -{1 \over 2} g_1 dp \big)^2 \bigg) \ .
\end{eqnarray}
Furthermore, the condition ({\ref{dccond}}) is equivalent to
\begin{eqnarray}
\label{dcond2}
d \psi &=& 4 \ell^{-2} e^x h^2(1-g_1 g_2) \psi_x dp \wedge dq
+ 2 \ell^{-1} g_1 h \psi_x dx \wedge dp
\nonumber \\
&+&\bigg(2 \ell^{-1} e^x g_2 h \psi_x -\psi_q \bigg) dx \wedge dq \ .
\end{eqnarray}

It remains to identify the basis $\{ V, W, \tau, \theta \}$ which is used to write the conditions in Section {\ref{conds}}.
We take
\begin{eqnarray}
W = dp
\end{eqnarray}
with
\begin{eqnarray}
\theta = {\ell^{-1} u \over \ell^{-2} u^2-h^2}(du+\psi)
+{\ell \over 2}{\ell^{-2} u^2+h^2 \over \ell^{-2} u^2 -h^2}
\bigg({1 \over 2} dx + \ell^{-1} e^x g_2 h dq + \ell^{-1} g_1 h dp \bigg) \ .
\end{eqnarray}
In particular, the above expression for $\theta$ is obtained directly from the gauge potential ({\ref{gtpot}}), on neglecting the $h^{-1}dh$ term arising in ${\cal{H}}$.
Then $\tau$ may be obtained by considering the condition
({\ref{dtxx}}), which for the Lorentzian Gauduchon-Tod solution in question can be written as
\begin{eqnarray}
\label{dtxx2}
\star d \theta &=& - d \tau + \ell^{-1} \theta \wedge \tau
+(V^1-V^3) \wedge \bigg(-h^{-1} d(h n_1)
\nonumber \\
&-&\ell^{-1}(g_1-\ell^{-1} h^{-1}u n_1)f h^{-1} 
\big(du+u({1 \over 2}dx +\ell^{-1} e^x g_2 h dq+\ell^{-1}g_1 h dp) + \psi\big)
\nonumber \\
&+&(\ell^{-1} h^{-1} u g_1 -n_1) V^2 - \ell^{-2} h^{-1} u (V^1+V^3) 
\nonumber \\
&+& \ell^{-1}h^{-1} (\psi_3-\psi_1) V^2 + {1 \over 2} \ell^{-1} h^{-1} \psi_2 (V^1+V^3) \bigg)
\end{eqnarray}
on setting $\psi = \psi_i V^i$, where we have set
\begin{eqnarray}
\tau &=& -f h^{-1}\big(du+u({1 \over 2}dx +\ell^{-1} e^x g_2 h dq+\ell^{-1}g_1 h dp) + \psi\big)
\nonumber \\
&+& \ell^{-1} h^{-1} u 
\bigg( {\ell \over 4} dx +{1 \over 2} e^x g_2 h dq -{1 \over 2}
g_1 h dp\bigg) + n_1 h dp
\end{eqnarray}
where $n_1=n_1(x,p,q)$ is a function. This function is required to satisfy the condition corresponding to the gauge choice $i_W \xi=0$, which can be read off from ({\ref{dtxx2}}) as
\begin{eqnarray}
\label{ngcond}
\nabla_W (h n_1) +2f \ell^{-2} \big(2  \psi_x- h^{-1} u \big) =0 \ .
\end{eqnarray}
Having fixed the three basis elements $W, \theta, \tau$, the remaining basis element $V$ must be given by
\begin{eqnarray}
V=h n_1 \tau + h g_1 \theta -{1 \over 2} h^2(n_1^2-g_1^2)dp 
-{1 \over 2}  e^x (\ell^{-2} u^2-h^2)dq \ .
\end{eqnarray}
It can then be checked directly that the basis
$\{ V, W, \tau, \theta \}$ defined in this way satisfies all of the conditions
in Section {\ref{conds}}, provided that the functions
$h, g_2, h_2, n_1$ satisfy the conditions ({\ref{lgte}}) and
({\ref{rxextra}}), and ({\ref{ngcond}}), and $\psi$ satisfies
({\ref{dcond2}}). For this solution, neither of $J^\pm$ are integrable, as ({\ref{icond}}) does not hold.

\subsection{Kastor-Traschen type solution}

A special case of the Gauduchon-Tod class of solutions for which the base space is
${\mathbb{R}}^{1,2}$ was also found in \cite{Klemm:2015mga}. Such solutions are a
split-signature analogue of the Kastor-Traschen solution \cite{Kastor:1992nn}.
The metric and gauge field strength are given by
\begin{eqnarray}
ds^2 = {\ell^2 \over (u+H)^2} du^2 + (u+H)^2\big(dz^2-dx^2-dy^2\big), \qquad A=-{\ell \over u+H} du
\end{eqnarray}
where $H(x,y,z)$ satisfies $(\partial_x^2+\partial_y^2-\partial_z^2)H=0$. For these solutions 
the basis $\{ V, W, \tau, \theta \}$ is given by
\begin{eqnarray}
W &=& dx+dz
\nonumber \\
\theta &=& {\ell \over u+H} du
\nonumber \\
\tau &=& (u+H)dy + \beta (dx+dz)
\nonumber \\
V &=& {1 \over 2} (u+H)^2 (dz-dx) + \beta(u+H) dy + {1 \over 2} \beta^2 (dx+dz)
\end{eqnarray}
with volume form ${\rm dvol}= \ell (u+H)^2 du \wedge dx \wedge dy \wedge dz$, and
$\beta=\beta(u,x,y,z)$ is a function which must satisfy
\begin{eqnarray}
{\partial \beta \over \partial z}-{\partial \beta \over \partial x}= {\partial H \over \partial y}
\end{eqnarray}
in order for the condition ({\ref{dtxx}}) to hold, with $\xi$ satisfying the gauge choice $i_W \xi=0$. With this choice, all of the geometric conditions listed in Section {\ref{conds}} hold.
Again, neither of $J^\pm$ are integrable in general.

\section{Example: Self-Dual Solutions}
Suppose that the gauge field strength is self-dual, $F=\star F$. Then the geometric conditions are
\begin{eqnarray}
\label{sdcond1}
d \theta = \star d \theta
\end{eqnarray}
and
\begin{eqnarray}
\label{sdcond2}
dW=0, \quad \nabla_W W=\nabla_W \tau=0, \quad \nabla_W V=-{1 \over \ell} \theta, \quad
\nabla_W \theta = {1 \over \ell} W
\end{eqnarray}
together with
\begin{eqnarray}
\label{sdcond3}
\star d \star \theta = -{3 \over \ell}, \quad \nabla_{\tau+\theta} \theta = {1 \over \ell} \tau, \quad \nabla_V W =  - i_{\theta +\tau}d \theta -{1 \over \ell} \theta
\end{eqnarray}
and there exists a 1-form $\xi$, satisfying $i_W \xi =i_V \xi=0$ such that
\begin{eqnarray}
\label{sdcond4}
d (\theta+\tau) = {1 \over \ell} \theta \wedge \tau -W \wedge \xi  \ .
\end{eqnarray}

We introduce local co-ordinates $u, v$ such that $W=du$, with dual tangent vector $W={\partial \over \partial v}$. On surfaces $u=const$, the condition ({\ref{sdcond4}}) implies
\begin{eqnarray}
({\tilde{\tau}}+{\tilde{\theta}}) \wedge d ({\tilde{\tau}}+{\tilde{\theta}}) =0
\end{eqnarray}
where ${\tilde{\tau}}$ and ${\tilde{\theta}}$ are the pull-backs of $\tau$, $\theta$ to surfaces of constant $u$.
It follows that there exists a local co-ordinate $p$, and functions ${\cal{G}}$, $H$ such that
\begin{eqnarray}
\label{hsocond}
\tau+\theta={\cal{G}} dp + H du .
\end{eqnarray}
We remark that the above geometric conditions imply that
\begin{eqnarray}
\label{sdcommut1}
[W, \theta+\tau] ={2 \over \ell}W
\end{eqnarray}
and hence a local co-ordinate $q$ can be found such that 
\begin{eqnarray}
\label{sdvect1}
\theta+\tau = {2v \over \ell}{\partial \over \partial v}+{\partial \over \partial q} \ .
\end{eqnarray}
The orthogonality conditions then imply that
\begin{eqnarray}
\label{sdbas1}
V=dv-{2 \over \ell}v dq +S du +h dp
\end{eqnarray}
for functions $S(u,v,p,q)$, $h(u,v,p,q)$, with
\begin{eqnarray}
\label{sdbas2}
\theta-\tau = m_1 du +m_2 dp +2 dq
\end{eqnarray}
for functions $m_1(u,v,p,q)$, $m_2(u,v,p,q)$. On substituting these basis 1-forms into the condition ({\ref{sdcond4}}), 
we find
\begin{eqnarray}
\partial_v {\cal{G}}=0, \qquad \partial_q {\cal{G}}={1 \over \ell} {\cal{G}}
\end{eqnarray}
which implies that ${\cal{G}}=e^{q \over \ell} \Psi$ for $\Psi=\Psi(u,p)$, and also
\begin{eqnarray}
\partial_v H=0
\end{eqnarray}
where the gauge condition $\xi_v=0$ has been used, so $H=H(u,p,q)$. These conditions imply that a basis transformation of the type ({\ref{basist1}}), where $\beta$ can be taken to be independent of the $v$ co-ordinate, together with a $\{u,p\}$ co-ordinate transformation, can be used to set, without loss of generality, $\Psi=1$ and $H=0$. Such a basis transformation  preserves the condition $\nabla_W \tau=0$, as well as the form of all the other geometric conditions.

It follows that the basis elements have been simplified to the following forms: 

\begin{eqnarray}
W &=& du
\nonumber \\
V &=& dv-{2v \over \ell} dq +S du +h dp
\nonumber \\
\theta &=& {1 \over 2}(m_2+e^{q \over \ell})dp +{1 \over 2}m_1 du +dq
\nonumber \\
\tau &=& {1 \over 2}(-m_2+e^{q \over \ell})dp -{1 \over 2}m_1 du -dq .
\end{eqnarray}
The condition ({\ref{sdcond4}}) holds with this basis. It remains to consider the remaining geometric conditions ({\ref{sdcond1}})-({\ref{sdcond3}}), with volume form
${\rm dvol}=e^{q \over \ell} du \wedge dv \wedge dp \wedge dq$.
After some calculation, the following conditions are obtained:

\begin{eqnarray}
\label{xxsd1}
\partial_u m_2 - \partial_p m_1 + \ell^{-1} e^{q \over \ell} m_1
+ \partial_v(h m_1 - S m_2)=0
\end{eqnarray}
\begin{eqnarray}
\label{xxsd2}
m_2 = -e^{q \over \ell} -\ell \partial_v h +{\ell \over 2} e^{q \over \ell} \partial_v m_1
\end{eqnarray}
\begin{eqnarray}
\label{xxsd3}
{1 \over 2} \ell^{-1} e^{q \over \ell} m_1 + \ell^{-1} v e^{q \over \ell} \partial_v m_1 -2 \ell^{-1} v \partial_v h
+{1 \over 2} e^{q \over \ell} \partial_q m_1 + 2 \ell^{-1} h - \partial_q h =0
\end{eqnarray}
\begin{eqnarray}
\label{xxsd4}
2 \ell^{-1} m_1 +2 \partial_v S +2 v \ell^{-1} \partial_v m_1 + \partial_q m_1 =0 .
\end{eqnarray}

To proceed to analyse these conditions, it will be convenient to set
\begin{eqnarray}
h = e^{q \over \ell} \big({1 \over 2} m_1 + \partial_v K \big)
\end{eqnarray}
for $K=K(u,v,p,q)$. Then the conditions ({\ref{xxsd2}}) and ({\ref{xxsd3}}) can be 
solved for $m_1$ and $m_2$ to give
\begin{eqnarray}
\label{mmsol}
m_1 = \partial_v \bigg( 2v \partial_v K -3K +\ell \partial_q K \bigg),
\qquad 
m_2 = -e^{q \over \ell} \bigg(1+\ell \partial_v^2 K \bigg)
\end{eqnarray}
and the condition ({\ref{xxsd4}}) can be integrated up to give
\begin{eqnarray}
\label{sfunct}
S = f(u,p,q) - 2 \ell^{-1} v^2 \partial_v^2 K -2v \partial_v \partial_q K
+ \ell^{-1} v \partial_v K +{3 \over 2} \partial_q K -{\ell \over 2} \partial_q^2 K
\end{eqnarray}
where the function $f$ is independent of $v$. In particular, this implies that
\begin{eqnarray}
h = e^{q \over \ell} \bigg(v \partial_v^2 K +{1 \over 2} \partial_v K + \ell \partial_v \partial_q K \bigg).
\end{eqnarray}
We remark that the function $f(u,p,q)$ appearing in $S$ can without loss of generality be set to zero by making an appropriate redefinition of $K$ as $K = \hat{K} + Z(u,p,q)$. This does not affect the form of the remaining conditions. Making this choice, the metric can be written entirely in terms of the function $K$ as
\begin{eqnarray}
\label{sdfinmet}
ds^2 &=& 2 du \bigg(dv 
+\big(-2 \ell^{-2} v^2 \partial_v^2 K -2v \partial_v \partial_q K + \ell^{-1} v \partial_v K
+{3 \over 2} \partial_q K -{\ell \over 2} \partial_q^2 K \big) du
\nonumber \\
&-&{2v \over \ell} dq + e^{q \over \ell}\big(2v \partial_v^2 K + \ell \partial_v \partial_q K \big) dp \bigg)
-e^{2q \over \ell} \big( 1+\ell \partial_v^2 K \big) dp^2 +2 e^{q \over \ell} dp dq.
\nonumber \\
\end{eqnarray}
This exhausts the content of the conditions ({\ref{xxsd2}}), ({\ref{xxsd3}}) and ({\ref{xxsd4}}).
It remains to consider ({\ref{xxsd1}}) which can be rewritten as
\begin{eqnarray}
\label{fsdc}
{\partial \over \partial v} \bigg[&& \hskip-6mm -2 \ell^{-1} e^{q \over \ell} v^2 \partial_v^2 K
-2 e^{q \over \ell} v \partial_v \partial_q K
+{3v \over \ell} e^{q \over \ell} \partial_v K
+{5 \over 2} e^{q \over \ell} \partial_q K -{\ell \over 2} e^{q \over \ell} \partial_q^2 K
\nonumber \\
&& \hskip-6mm  +3 \partial_p K -\ell \partial_p \partial_q K
-\ell e^{q \over \ell} \partial_v \partial_u K -2v \partial_v \partial_p K -3 \ell^{-1} e^{q \over \ell} K -{1 \over 2} e^{q \over \ell} (\partial_v K)^2
\nonumber \\
&&\hskip-6mm  +{\ell^2 \over 2} e^{q \over \ell} (\partial_v \partial_q K)^2
+ e^{q \over \ell} v \partial_v K \partial_v^2 K +{3 \over 2} \ell e^{q \over \ell} \partial_q K \partial_v^2 K -{\ell^2 \over 2} e^{q \over \ell} \partial_q^2 K \partial_v^2 K \bigg] =0.
\nonumber \\
\end{eqnarray}
In particular, the metric ({\ref{sdfinmet}}) together with the condition
({\ref{fsdc}}) automatically satisfies the Einstein condition
\begin{eqnarray}
R_{\mu \nu} = -{3 \over \ell^2} g_{\mu \nu}
\end{eqnarray}
with the exception of the co-ordinate indices $\mu=\nu=u$, as is expected from consideration of the integrability conditions obtained from the Killing spinor equation, which were considered
in \cite{Gutowski:2019hnl}. Imposing the $\mu = \nu=u$ component
of this equation produces a further nonlinear PDE in the function $K$ which is not implied
by ({\ref{fsdc}}). 

\section{Conclusions}

We have determined the conditions on the geometry for a solution of minimal gauged supergravity with positive cosmological constant in neutral signature to preserve the minimal $N=1$ amount of supersymmetry. This class of solutions was omitted from the classification constructed in
\cite{Klemm:2015mga}. It would be interesting to understand this geometric structure better. Furthermore, we have shown how the $N=2$ solutions which are a fibration over a 3-dimensional Lorentzian Gauduchon-Tod base space arise in this structure.

Gauduchon-Tod metrics have arisen in a number of different ways
in the context of $D=4$ supergravity.
Euclidean Gauduchon-Tod structures have been found in
minimal Euclidean $D=4$ supergravity with signature $(+,+,+,+)$ \cite{euclidean3}, and also in minimal $D=4$ de Sitter supergravity with
signature $(-,+,+,+)$ \cite{Gutowski:2009vb}. In the former case, there are no Majorana spinors. In the latter case, there exists a charge conjugation operator $C*$ which commutes with the gamma matrices, however it does not commute with the supercovariant derivative. This means that we do not expect
that these Euclidean Gauduchon-Tod solutions can be viewed
as $N=2$ supersymmetric solutions which are special cases
of a more general $N=1$ geometric structure, as is the case for 
the solutions considered in this paper.

In contrast, Lorentzian Gauduchon-Tod structures were also found in minimal gauged $D=4$ pseudo-supergravity \cite{Gutowski:2018shj}, with signature $(-,+,+,+)$.  
In this case, the charge conjugation operator commutes with
the supercovariant derivative, and we expect that the
Lorentzian Gauduchon-Tod solution found in that case can also
be written as a special case of the supersymmetric Majorana solutions which were also classified in section 4.2 of \cite{Gutowski:2018shj}, on making an appropriate choice of co-ordinates.

\newpage

\setcounter{section}{0}
\setcounter{subsection}{0}

\appendix{The linear system}

In this appendix we present the linear system which is equivalent to the KSE ({\ref{kse1}}) in the case for which
the Majorana Killing spinor is non-chiral. In particular,
following the conventions of \cite{Gutowski:2019hnl},
with a split signature (pseudo)-holomorphic basis,
i.e. a basis 
\begin{eqnarray}
{\bf{e}}^1, \quad {\bf{e}}^2, \quad {\bf{e}}^{\bar{1}}=({\bf{e}}^1)^*,
\quad {\bf{e}}^{\bar{2}}=({\bf{e}}^2)^*
\end{eqnarray}
with respect to which the metric is
\begin{eqnarray}
\label{cframe}
ds^2 = 2 {\bf{e}}^1 {\bf{e}}^{\bar{1}}
-2 {\bf{e}}^2 {\bf{e}}^{\bar{2}} \ ,
\end{eqnarray}
the linear system obtained from ({\ref{kse1}}) is as follows:
\begin{eqnarray}
\label{linsys}
-\omega_{1,1\bar{1}}+\omega_{1,2\bar{2}}+2i \omega_{1,\bar{1} \bar{2}}&=&\sqrt{2}(-F_{1 \bar{1}}+F_{2 \bar{2}}) +{2 \over \ell} A_1 +{\sqrt{2} \over \ell}
\nonumber \\
\omega_{1,1 \bar{1}}-\omega_{1,2 \bar{2}}-2i \omega_{1,12}&=&-2 \sqrt{2}i F_{12}
+{2 \over \ell}A_1
\nonumber \\
\omega_{1,1 \bar{1}}+\omega_{1,2 \bar{2}}-2i \omega_{1,1 \bar{2}}&=& -2 \sqrt{2}i F_{1 \bar{2}} +{2 \over \ell}A_1
\nonumber \\
- \omega_{1,1 \bar{1}}-\omega_{1,2 \bar{2}}+2i \omega_{1, \bar{1} 2} &=&-\sqrt{2}(F_{1 \bar{1}}+F_{2 \bar{2}}) +{2 \over \ell}A_1 +{\sqrt{2} \over \ell}
\nonumber \\
-\omega_{2,1\bar{1}}+\omega_{2,2\bar{2}}+2i \omega_{2,\bar{1} \bar{2}}&=&\sqrt{2}i(-F_{1 \bar{1}}+F_{2 \bar{2}})+{2 \over \ell}A_2 +{\sqrt{2} i \over \ell}
\nonumber \\
\omega_{2,1 \bar{1}}-\omega_{2,2 \bar{2}}-2i \omega_{2,12}&=&2 \sqrt{2} F_{12}+{2 \over \ell}A_2
\nonumber \\
\omega_{2,1 \bar{1}}+\omega_{2,2 \bar{2}}-2i \omega_{2,1 \bar{2}}&=& -\sqrt{2}i (F_{1 \bar{1}}+F_{2 \bar{2}}) +{2 \over \ell}A_2 -{\sqrt{2} i \over \ell}
\nonumber \\
- \omega_{2,1 \bar{1}}-\omega_{2,2 \bar{2}}+2i \omega_{2, \bar{1} 2} &=&2 \sqrt{2} F_{\bar{1} 2} +{2 \over \ell}A_2 \ .
\end{eqnarray}
The spinor bilinears ({\ref{bilin}}) are given by
\begin{eqnarray}
\label{cc1}
W = 2 \sqrt{2}i ({\bf{e}}^1-{\bf{e}}^{\bar{1}})-2 \sqrt{2}({\bf{e}}^2+{\bf{e}}^{\bar{2}})
\end{eqnarray}
and
\begin{eqnarray}
\label{cc2}
\chi =W \wedge \theta, \qquad \theta = {1 \over \sqrt{2}}({\bf{e}}^1 + {\bf{e}}^{\bar{1}}) \ .
\end{eqnarray}
The conditions ({\ref{cc3}}) and ({\ref{cc4}}), with the bilinears ({\ref{cc1}}) and ({\ref{cc2}}), are equivalent to the linear
system ({\ref{linsys}}).

{\flushleft{\textbf{Acknowledgements:}}} JG is supported by the STFC
Consolidated Grant ST/L000490/1. JG would like to thank
the Department of Mathematics, University of Liverpool, 
for hospitality during which part of this work was completed.
The work of WS is supported in part by
the National Science Foundation under grant number PHY-1620505. The authors would like to thank
Maciej Dunajski for useful conversations.

\end{document}